\documentstyle[12pt,epsfig,aaspp]{article}
\lefthead{Gallart et al.}
\righthead{{\it HST} observations of Leo I}

\begin{document}

\title{{\it HST} observations of the Local Group dwarf galaxy Leo~I}

\author{Carme Gallart\altaffilmark{1}, Wendy L. Freedman\altaffilmark{1}, Mario Mateo\altaffilmark{2}, Cesare Chiosi\altaffilmark{3}, Ian B. Thompson\altaffilmark{1}, Antonio Aparicio\altaffilmark{4}, Giampaolo Bertelli\altaffilmark{3, 5}, Paul W. Hodge\altaffilmark{6}, Myung G. Lee\altaffilmark{7}, Edward W. Olszewski\altaffilmark{8}, Abhijit Saha\altaffilmark{9}, Peter B. Stetson\altaffilmark{10}, Nicholas B. Suntzeff\altaffilmark{11}}

Subject Headings:  galaxies: individual (Leo I); galaxies: evolution; galaxies: stellar content; galaxies: photometry; stars: Hertzsprung-Russell (HR-diagram).

\altaffiltext{1}{Observatories of the Carnegie Institution of Washington, 813 Santa Barbara St., Pasadena, California 91101, USA} 
\altaffiltext{2}{Department of Astronomy, 821 Dennison Building, University of Michigan, Ann Arbor, Michigan 48109, USA}
\altaffiltext{3}{Dipartimento di Astronomia dell'Universit\`a di Padova, Vicolo dell'Osservatorio 5, I-35122- Padova, Italy}
\altaffiltext{4}{Instituto de Astrof\'\i sica de Canarias, E-38200 La Laguna, Canary Islands, Spain} 
\altaffiltext{5}{National Council of Research, CNR-GNA, Rome, Italy}
\altaffiltext{6}{Department of Astronomy, University of Washington, Seattle, WA 98195, USA}
\altaffiltext{7}{Department of Astronomy, Seoul National University, Korea}
\altaffiltext{8}{Steward Observatory, The University of Arizona, Tucson, AZ 85721, USA}
\altaffiltext{9}{National Optical Astronomy Observatory, P.O. Box 26732, Tucson, AZ 85726, USA}
\altaffiltext{10}{National Research Council, Hertzberg Institute of Astrophysics, Dominion Astrophysical Observatory, Canada}
\altaffiltext{11}{Cerro Tololo Inter--American Observatory, National Optical Astronomy Observatory, Casilla 603, La Serena, Chile }

\begin{abstract}

We present deep $HST$ F555W ($V$) and F814W ($I$) observations of a
central field in the Local Group dwarf spheroidal (dSph) galaxy Leo~I.
The resulting color-magnitude diagram (CMD) reaches $I \simeq 26$ and
reveals the oldest  $\simeq 10-15$ Gyr old turnoffs. Nevertheless, a
horizontal branch is not obvious in the CMD. Given the low metallicity of the galaxy, this likely indicates that the first
substantial star formation in the galaxy may have been somehow delayed
in Leo~I in comparison with the other dSph satellites of the Milky
Way. The subgiant region is well and uniformly populated from the
oldest turnoffs up to the 1 Gyr old turnoff, indicating that
star formation has proceeded in a continuous way, with possible 
variations in intensity but no big gaps between successive bursts, over the
galaxy's lifetime. The structure of the red-clump of core
He-burning stars is consistent with the large amount of
intermediate--age population inferred from the main sequence and the
subgiant region. In spite of the lack of gas in Leo~I, the CMD clearly shows star formation continuing until 1 Gyr ago and possibly until a few hundred
Myrs ago in the central part of the galaxy.

\end{abstract}

\section{Introduction} \label{intro}

The first color-magnitude diagrams (CMD) obtained by Baade for the
dwarf spheroidal (dSph) companions of the Milky Way, and in particular for
the Draco system (Baade \& Swope 1961), showed all of the features
present in the CMD's of globular clusters. This, together with the presence of RR
Lyrae stars (Baade \& Hubble 1939; Baade \& Swope 1961) led to
the interpretation that dSph galaxies are essentially pure Population II
systems. But Baade (1963) noted that there are a number of
characteristics in the stellar populations of dSph galaxies that
differentiate them from globular clusters, including extreme red 
horizontal branches and the distinct
characteristics of the variable stars. When carbon stars were
discovered in dSph galaxies,  these
differences were recognized to be due to the presence of an
intermediate-age population (Cannon, Niss \& Norgaard--Nielsen 1980;
Aaronson, Olszewski \& Hodge 1983; Mould \& Aaronson 1983). In the past few years this intermediate-age population has been shown beautifully in the 
CMDs of a number of dSph galaxies (Carina: Mould \& Aaronson 1983; Mighell 1990; Smecker-Hane, Stetson \& Hesser 1996; Hurley-Keller, Mateo \& Nemec 1998; Fornax: Stetson, Hesser \& Smecker-Hane 1998; Leo I: Lee et al. 1993, L93 hereinafter; this paper). Other dSph show only a dominant old stellar population in their CMDs (Ursa Minor: Olszewski \& Aaronson 1985; Mart\'\i nez-Delgado \& Aparicio 1999; Draco: Carney \& Seitzer 1986; Stetson, VandenBergh \& McClure 1985; Grillmair et al. 1998; Sextans: Mateo et al. 1991).

An old stellar population, traced by a horizontal-branch (HB), has been
clearly observed in all the dSph galaxies satellites of the Milky Way, except Leo~I, regardless of their subsequent star formation histories (SFH). In this respect, as noted by L93, Leo~I is a peculiar galaxy, showing a well populated red-clump (RC) but no evident HB. This suggests that the first substantial amount of star formation may have been somehow delayed in this galaxy compared with the other dSph. Leo~I is also singular in that its large galactocentric radial velocity (177$\pm$3 km $\rm {s}^{-1}$, Zaritsky et al. 1989) suggests that it may not be bound to the Milky Way, as the other dSph galaxies seem to be (Fich \& Tremaine 1991). Byrd et al. (1994) suggest that both Leo~I and the Magellanic Clouds seem to have left the neighborhood of the
Andromeda galaxy about 10 Gyr ago. It is interesting that the Magellanic Clouds also seem to have only a small fraction of old stellar population.

Leo~I presents an enigmatic system with unique characteristics among
Local Group galaxies. From its morphology and from its similarity to other dSph in terms of its lack of detectable quantities of HI (Knapp, Kerr \& Bowers 1978, see Section~\ref{leoi_prev}) it would be considered a dSph galaxy. But it
also lacks a conspicuous old population and it has a much larger fraction of
intermediate-age population than its dSph counterparts, and even, a
non-negligible population of young ($\le$ 1 Gyr old) stars.

In this paper, we present new {\it HST} F555W ($V$) and F814W ($I$)
observations of Leo~I.  In Section~\ref{leoi_prev}, the previous work
on Leo~I is briefly reviewed. In Section~\ref{obs}, we present the
observations and data reduction. In Section~\ref{phot} we discuss the
photometry of the galaxy, reduced independently using both ALLFRAME and
DoPHOT programs, and calibrated using the ground-based photometry of L93. In Section~\ref{cmd} we present the CMD  of Leo~I, and discuss the stellar populations and the metallicity of the galaxy. In Section~\ref{discus} we summarize the conclusions of this paper. In a companion paper, (Gallart et al. 1998, Paper~II) we will quantitatively derive the SFH of Leo~I through the comparison of the observed CMD with a set of synthetic CMDs.

\section{Previous Work on Leo~I} \label{leoi_prev}

Leo~I (DDO~74), together with Leo~II, was discovered by Harrington \&
Wilson (1950) during the course of the first Palomar Sky Survey. The
distances to these galaxies were estimated to be $\simeq$ 200 kpc, considerably
more distant than the other dSph companions of the Milky Way.

It has been observed in HI by Knapp et al. (1978) using the NRAO 91-m telescope, but not detected. They set a limit for its HI mass of $M_{HI}/M_{\odot} \le 7.2\times10^3$ in the central 10\arcmin ($\simeq$ 780 pc) of the galaxy. Recently, Bowen et al. (1997) used spectra of three QSO/AGN to set a limit on the HI column density within 2--4 kpc in the halo of Leo~I to be $N(HI) \le 10^{17} {\rm cm}^{-2}$. They find no evidence of dense flows of gas in or out of Leo~I, and no evidence for tidally disrupted gas. 

The large distance to Leo I and the proximity on the sky of the bright
star Regulus have made photometric studies difficult.  As a consequence, 
the first CMDs of Leo~I were obtained much later than for the other nearby dSphs (Fox \& Pritchet 1987; Reid \& Mould 1991; Demers, Irwin \& Gambu 1994; L93). From the earliest observations of the stellar populations
of Leo I there have been indications of a large quantity of
intermediate-age stars.  Hodge \& Wright (1978) observed an unusually
large number of anomalous Cepheids, and carbon stars were found by
Aaronson et al. (1983) and Azzopardi, Lequeux \& Westerlund (1985,
1986). A prominent RC, indicative in a low Z system of an intermediate-age stellar population, is seen both in the $[(B-V, V]$ CMD of Demers et al. (1994)
and in the $[(V-I), I]$ CMD of L93. The last CMD is
particularly deep, reaching $I\simeq 24$ ($M_I \simeq + 2$), and
suggests the presence of a large number of intermediate--age,
main--sequence stars. There is no evidence for a prominent HB in any of
the published CMD's.

L93 estimated the distance of Leo~I to be $(m-M)_0 = 22.18 \pm 0.11$ based on the position of the tip of the red giant
branch (RGB); we will adopt this value in this paper. They also
estimated  a metallicity of [Fe/H] = --2.0$\pm$0.1 dex  from the mean
color of the RGB. Previous estimates of the
metallicity (Aaronson \& Mould 1985; Suntzeff, Aaronson \& Olszewski
1986; Fox \& Pritchet 1987; Reid \& Mould 1991) using a number of
different methods  range from [Fe/H]=--1.0 to --1.9 dex. With the new {\it HST} data presented in this paper, the information on the age structure from the turnoffs will help to further constrain the metallicity.

\section{Observations and Data Reduction} \label{obs}

We present WFPC2 {\it HST} $V$ (F555W) and $I$ (F814W) data in one
2.6\arcmin $\times$ 2.6\arcmin~field in Leo~I obtained in March 5,
1994. The WFPC2 has four internal cameras: the planetary camera (PC)
and three Wide Field (WF) cameras. They image onto a Loral
800$\times$800 CCD, which gives an scale of 0\arcsec.046 pixel$^{-1}$
for the PC camera and 0\arcsec.10 pixel$^{-1}$ for the WF cameras. At
the time of the observations the camera was still operating at the higher
temperature of --77.0 $^o$C. Figure~\ref{carta} shows the location of
the WFPC2 field superimposed on Digitized Sky Survey image of Leo~I.
The position of the ground-based image of L93 is also
shown.   The position was chosen so that the PC field was situated in the central,
more crowded part of the galaxy.  Three deep exposures in both F555W
($V$) and F814W ($I$) filters (1900 sec. and 1600 sec. each,
respectively) were taken. To ensure that the brightest stars were not saturated,  one shallow exposure in each filter (350 sec. in F555W and 300 sec in F814W) was also obtained. Figure~\ref{mosaic} shows the $V$ and $I$ deep (5700 sec. and 4800 sec. respectively) WF Chip2 images of Leo~I.

\begin{figure}
\caption[]{Digitized Sky Survey image of the Leo~I field. The outlines indicate the WFPC2 field and the field observed by L93. The total field shown here is $10\arcmin\times 10\arcmin$. North is up, east is to the left.}
\label{carta}
\end{figure}

\begin{figure}
\caption[]{$V$ (above) and $I$ (below) deep (5700 sec. and 4800 sec. respectively)  WF Chip~2 Leo~I images. }
\label{mosaic}
\end{figure}

All observations were preprocessed through the standard STScI pipeline,
as described by Holtzmann et al. (1995). In addition, the treatment of
the vignetted edges, bad columns and pixels, and correction of the
effects of the geometric distortion produced by the WFPC2 cameras, were
performed as described by Silbermann et al. (1996). 


\section{Photometry} \label{phot}
\subsection{Profile fitting photometry}  \label{psfphot}

Photometry of the stars in Leo~I was measured independently using the set
of DAOPHOT~II/ALLFRAME programs  developed by Stetson (1987, 1994),
and also with a  modified version of DoPHOT (Schechter, Mateo \& Saha
1993). We compare the results obtained with each of these programs below.

ALLFRAME photometry was performed in the 8 individual frames and the photometry list in each band was obtained by averaging the magnitudes of the corresponding individual frames. In summary, the process is as follows: a candidate star list
was obtained from the median of all the images of each field using
three DAOPHOT~II/ALLSTAR detection passes. This list was fed to
ALLFRAME, which was run on all eight individual frames simultaneously.
We have used the PSFs obtained from the public domain
{\it HST} WFPC2 observations of the globular clusters Pal 4 and
NGC~2419 (Hill et al. 1998). The stars in the different frames of
each band were matched and retained if they were found in at least three
frames for each of $V$ and $I$. The magnitude of each star in each band
was set to the error-weighted average of the magnitudes for each star
in the different frames. The magnitudes of the brightest stars were
measured from the short exposure frames. A last match between the stars
retained in each band was made to obtain the $VI$ photometry table.

DoPHOT photometry was obtained with a modified version of the code to account for the {\it HST} PSF  (Saha et al. 1996). DoPHOT reductions were made on  average $V$ and $I$ images combined in a manner similar to that described by Saha et al. (1994) in order to remove the effects of cosmic rays.  Photometry of the brightest stars was measured  from the $V$ and $I$ short exposure frames.


The DoPHOT and ALLFRAME calibrated photometries (see Section~\ref{transjohn}) show a reasonably good agreement.  There is a scatter of 2-3\% for
even the brightest stars in both $V$ and $I$. No systematic differences can be seen in the $V$ photometry. In the $I$ photometry there is good systematic agreement among the brightest stars, but a small tendency for the DoPHOT magnitudes to become brighter compared to the ALLFRAME magnitudes with increasing $I$ magnitude.  This latter effect is about 0.02 magnitudes at the level of the RC, and increases to about 0.04-0.05 mag by $I$ = 26.  We cannot decide from these data which program is `correct'. However, the systematic differences are sufficiently small compared to the random scatter that our final conclusions are identical regardless of which reduction program is used.

In the following we will use the star list obtained with
DAOPHOT/ALLFRAME. Our final photometry table contains a total of 31200
stars found in the four WFPC2 chips after removing stars with
excessively large photometric errors compared to other stars of similar
brightness. The retained stars have $\sigma\le 0.2$, chi$<1.6$ and $-0.5\le$sharp$\le 0.5$.

\subsection{Transformation to the Johnson-Cousins system} \label{transjohn}

For our final photometry in the Johnson-Cousins system we will rely ultimately in the photometry obtained by L93. Before this last step though, we transformed the profile fitting photometry using the prescription of Holtzmann et al. (1995). In this section, we will describe both steps and discuss the differences between the {\it HST}-based photometry and the ground based photometry.    

The ALLFRAME photometry has been transformed to standard magnitudes in the Johnson-Cousins system using  the prescriptions of Holtzmann et al. (1995) and Hill et al. (1998) as adopted for the {\it HST} $H_0$ Key Project data. PSF magnitudes have been transformed to instrumental magnitudes at an aperture of radius 0.5" (consistent with Holtzmann et al. 1995 and Hill et al. 1998) by deriving the value for the aperture correction for each frame using DAOGROW (Stetson 1990). 

The Johnson-Cousins magnitudes obtained in this way were compared with ground-based magnitudes for the same field obtained by L93 by matching a number of bright ($V<21, I<20$), well measured stars in the {\it HST}
($V_{HST}$, $I_{HST}$) and ground-based photometry ($V_{Lee}$,
$I_{Lee}$). The zero-points between both data sets have been determined
as the median of the distribution of ($V_{Lee}-V_{HST}$) and
($I_{Lee}-I_{HST}$). In Table~\ref{zeros} the values for the median of ($V_{Lee}-V_{HST}$), ($I_{Lee}-I_{HST}$) and its dispersion $\sigma$ are listed for each chip (no obvious color terms are observed, as expected, since both photometry sets have been transformed to a standard system taking into account the color terms where needed of the corresponding telescope-instrument system). $N$ is the number of stars used to calculate the transformation. Although the value of the median zero--point varies from chip to chip, it is in the sense of making the corrected $V$ magnitudes brighter by $\simeq 0.05$ mag than $V_{HST}$ and the corrected $I$ magnitudes fainter than $I_{HST}$ by about the same amount. Therefore, the final $(V-I)$ colors are one tenth of a magnitude bluer in the corrected photometry.

\begin{table}
\caption {Zero points: ($V_{Lee}-V_{HST}), (I_{Lee}-I_{HST})$}
\label{zeros} 
\begin{center}
\begin{tabular}{lcccc}
\hline
\hline
\noalign{\vspace{0.1 truecm}}
CHIP & filter &  median & $\sigma$& N \\
\noalign{\vspace{0.1 truecm}}
\hline
\noalign{\vspace{0.1 truecm}}
CHIP 1 &  F555W  & -0.037 & 0.100   & 17 \\ 
CHIP 2 &  F555W  & -0.110 & 0.103   & 59 \\
CHIP 3 &  F555W  & -0.080 & 0.063   & 43 \\
CHIP 4 &  F555W  & -0.059 & 0.067   & 57 \\ 
\noalign{\vspace{0.1 truecm}}
\hline
CHIP 1 &  F814W  & 0.035 & 0.075   &  17 \\ 
CHIP 2 &  F814W  & 0.013 & 0.064   &  53 \\
CHIP 3 &  F814W  & 0.076 & 0.042   &  43 \\
CHIP 4 &  F814W  & 0.080 & 0.041   &  53 \\
\noalign{\vspace{0.1 truecm}}
\hline
\hline 
\end{tabular}
\end{center}
\end{table}

Note that the CTE effect, which may be important in the case of observations made at the temperature of $-77^o$ C, could contribute to the dispersion on the zero-point. Nevertheless, if the differences $(V_{Lee}-V_{HST})$, $(I_{Lee}-I_{HST})$ are plotted for different row intervals, no clear trend is seen, which indicates that the error introduced by the CTE effect is not of concern in this case. The fact that the background of our images is considerable (about 70 $e^-$) can be the reason for the CTE effect not being noticeable.

We adopt the L93  calibration because it was based on observations of a large number of standards from Graham (1981) and Landolt (1983) and because there was very good agreement between independent calibrations performed on two different observing runs and between calibrations on four nights of one of the
runs. In addition, the Holtzmann et al. (1995) zero points were derived
for data taken with the Wide Field Camera CCD's operating at a lower temperature compared to the present data set.

\begin{figure}
\caption[]{Observed CMDs of Leo~I for all four WFPC2 chips. }
\label{4cmd}
\end{figure}

\section{The Leo~I color-magnitude diagram}\label{cmd}

\subsection{Overview}\label{overview}

 In Figure~\ref{4cmd} we present four $[(V-I), I]$ CMDs for Leo~I based
on the four WFPC2 chips. Leo~I possesses a rather steep and blue RGB, 
indicative of a low metallicity. Given this low metallicity, its very well-defined RC, at $I\simeq$ 21.5, is characteristic of an
intermediate-age stellar population. The main sequence (MS), reaching
up to within 1 mag in brightness of the RC, unambiguously shows that a
considerable number of stars with ages between $\simeq$ 1 Gyr and 5 Gyr
are present in the galaxy, confirming the suggestion by L93
that the faintest stars in their photometry might be from a
relatively young ($\simeq$ 3 Gyr) intermediate-age population. Our CMD,
extending about 2 magnitudes deeper than the L93 photometry and
reaching the position expected for the turnoffs of an old population, 
shows that a rather broad range in ages is present in Leo~I. A number of yellow stars, slightly brighter and bluer than the RC, are probably evolved counterparts of the brightest stars in the MS. Finally, the lack of discontinuities in the turnoffs/subgiant region indicate a continuous star formation activity (with possible changes of the star formation rate intensity) during the galaxy's lifetime.

We describe each of these features in more detail in Section~\ref{compaiso}, and discuss their characteristics by comparing them with theoretical isochrones 
and taking into account the errors discussed in Section~\ref{photerrors}. We
will quantitatively study the SFH of Leo~I in Paper~II by comparing the distribution of stars in the observed CMD with a set of model CMDs computed using the stellar evolutionary theory as well as a realistic simulation of the observational effects in the photometry (see Gallart et al. 1996b,c and Aparicio, Gallart \& Bertelli 1997 a,b for different applications of this method to the study of the SFH in several LG dwarf irregular galaxies).

\subsection{Photometric errors}\label{photerrors}

Before proceeding with an interpretation of the features present in the CMD, it is important to assess the photometric errors. To investigate the total errors present in the photometry, artificial star tests have been performed in a similar way as described in Aparicio \& Gallart (1994) and Gallart, Aparicio \& V\'\i lchez (1996a). For details on the tests run for the Leo~I data, see Paper II.  In short, a large number of artificial stars of known magnitudes and colors were injected into the original frames, and the photometry was redone again following exactly the same procedure used to obtain the photometry for the original frames. The injected and recovered magnitudes of the artificial stars, together with the information of the artificial stars that have been lost, provides us with the true total errors.

In Figure~\ref{errors}, artificial stars representing a number of small intervals of magnitude and color have been superimposed as white spots on the observed CMD of Leo~I. Enlarged symbols ($\times$, $\circ$, $\triangle$) show the recovered magnitudes for the same artificial stars. The spread in magnitude and color shows the error interval in each of the selected positions. This information will help us in the interpretation of the different features present in the CMD (Section~\ref{compaiso}). A more quantitative description of these errors and a discussion of the characteristics of the error distribution will be presented in Appendix A of Paper~II.

\begin{figure}
\caption[]{A fraction of the results of the artificial star test conducted on the Leo~I image, superimposed in selected positions of the observed CMD of Leo~I. White spots show the locus of the injected magnitudes of a set of artificial stars, and the enlarged symbols show the recovered magnitudes for the same artificial stars. The scatter in the recovered magnitudes gives us information on the total errors in each position.}
\label{errors}
\end{figure}

\subsection{The Leo I distance}

We will adopt, here and in Paper II the distance obtained by L93 $(m-M)_0=22.18\pm0.11$ from the position of the tip of the RGB. Since the ground based observations of L93 cover a larger area than the {\it HST} observations presented in this paper, and therefore sample the tip of the RGB better, they are more suitable to derive the position of the tip. On another hand, since we derive the callibration of our photometry from theirs, we don't expect any difference in the position of the tip in our data. The adopted distance provides a good agreement between the position of the different features in the CMD and the corresponding theoretical position (Figures~\ref{leoi_isopa} and~\ref{leoi_isoya}), and its uncertainty does not affect the (mostly qualitative) conclusions of this paper.      

\subsection{Discussion of the CMD of Leo~I: Comparison with theoretical isochrones}\label{compaiso}

\begin{figure}
\caption[]{Combined $[(V-I)_0, M_I]$ CMD of the stars in the WFPC2 field. A distance modulus $(m-M)_0=22.18$ (L93) and reddening E(B-V)=0.02 (Burstein \& Heiles 1984) have been used to transform to absolute magnitudes and unreddened colors. Isochrones of 16 Gyr (Z=0.0004) --thick line--, 3 Gyr (Z=0.001) --thin line-- and 1 Gyr  (Z=0.001) --dashed line-- from the Padova library (Bertelli et al. 1994) have been superimposed on the data. Only the evolution to the tip of the RGB is shown in the 16 Gyr and 3 Gyr isochrones. The Z=0.001 isochrones published by Bertelli et al. (1994) are calculated using the old Los Alamos opacities (Huebner et al. 1977), and therefore, are not homogeneous with the rest of the isochrones of their set. The Z=0.001 isochrones drawn here have been calculated by interpolation between the Z=0.0004 and Z=0.004 isochrones.}
\label{leoi_isopa}
\end{figure}

\begin{figure}
\caption[]{Yale isochrones (Demarque et al. 1996) for the same ages, metallicities and evolutionary phases (except for the 1 Gyr old isochrone) as in Figure~5, superimposed on the same data.} 
\label{leoi_isoya}
\end{figure}

\begin{figure}
\caption[]{HB-AGB phases for 16 Gyr ($Z=0.0004$) --thick line--  and complete isochrones of 1 Gyr, 600 and 400 Myr ($Z=0.001$) --thin lines-- from the Padova library, superimposed on the same data of Figure~5. See details on the Z=0.001 isochrones in the caption of Figure~5.}
\label{leoi_isoclump}
\end{figure}

In Figure~\ref{leoi_isopa}, isochrones of 16 Gyr ($Z=0.0004$), 3 and 1 Gyr ($Z=0.001$) from the Padova library (Bertelli et al. 1994) have been superimposed upon the global CMD of Leo~I. In Figure~\ref{leoi_isoya}, isochrones of the same ages and metallicities from the Yale library (Demarque et al. 1996) are shown. In both cases (except for the Padova 1 Gyr old,  Z=0.001 isochrone), only the evolution through the RGB tip has been displayed (these are the only phases available in the Yale isochrones). In Figure~\ref{leoi_isoclump}, the HB--AGB phase for 16 Gyr ($Z=0.0004$) and the full isochrones for 1 Gyr, 600 and 400 Myr ($Z=0.001$) from the Padova library are shown.  

A comparison of the Yale and Padova isochrones in
Figures~\ref{leoi_isopa} and ~\ref{leoi_isoya} shows some differences
between them, particularly regarding the shape of the RGB (the RGB of
the Padova isochrones are in general {\it steeper} and redder, at the
base of the RGB, and bluer near the tip of the RGB, than the Yale 
isochrones for the same age and Z) and
the position of the subgiant branches of age $\simeq 1$Gyr (which is
brighter in the Padova isochrones).  In spite of these
differences, the general characteristics deduced for the stellar
populations of Leo~I do not critically depend on the set chosen. 
However, based on these comparisons, we can gain some insight into current
discrepancies between two sets of evolutionary models widely used, and
therefore into the main uncertainties of stellar evolution theory that
we will need to take into account when analyzing the observations using
synthetic CMDs (Paper~II).

In the following, we will discuss the main features of the Leo~I CMD using the isochrones in Figures~\ref{leoi_isopa} to~\ref{leoi_isoz}. This will allow us to reach a qualitative understanding of the stellar populations of Leo~I, as a starting point of the more quantitative approach presented in Paper~II. 

\subsubsection{The main-sequence turnoff/subgiant region}\label{ms}

The broad range in magnitude in the MS turnoff region of Leo~I CMD is a clear indication of a large range in the age of the stars populating Leo~I. The fainter envelope of the subgiants  coincides well with the position expected for a $\simeq$ 10--15 Gyr old population whereas the brightest blue stars on the main sequence (MS) may be as young as 1 Gyr old, and possibly
younger. 

Figure~\ref{leoi_isoclump} shows that the blue stars brighter than the 1 Gyr isochrone are well matched by the MS turnoffs of stars a few hundred Myr old. One may argue that a number of these stars may be affected by observational errors that, as we see in Figure~\ref{errors}, tend to make stars brighter.  They could also be unresolved binaries comprised of two blue stars. Nevertheless, it is very unlikely that the brightest blue stars are stars $\simeq$ 1 Gyr old affected by one of these situations, since one has to take into account that: a) a 1 Gyr old binary could be only as bright as $M_I\simeq 0.3$ in the extreme case of two identical stars, and b) none of the blue artificial stars at $M_I\simeq 1$ (which are around 1 Gyr old) got shifted the necessary amount to account for the stars at $M_I \simeq -0.1$ and only about 4\% of them have been shifted a maximum of 0.5 mag. We conclude, therefore, that some star formation has likely been going on in the galaxy from 1 Gyr to a few hundreds Myr ago. The presence of the bright yellow stars (see subsection~\ref{yellow} below), also supports this conclusion. 

Concerning the age of the older population of Leo~I, the present analysis of the  data using isochrones alone does not allow us to be much more precise than the range given above (10--15 Gyr), although  we favour the hypothesis that there may be stars older than 10 Gyr in Leo~I. In the old age range, the isochrones are very close to one another in the CMD and therefore the age resolution is not high. In addition, at the corresponding magnitude, the observational errors are quite large. Nevertheless, the characteristics of the errors as shown in Figure~\ref{errors} make it unlikely that the faintest stars in the turnoff region are put there due to large errors because i) a significant migration to fainter magnitudes of the stars in the $\simeq$10 Gyr turnoff area is not expected and, ii) because of the approximate symmetric error distribution, errors affecting intermediate-age stars in their turnoff region are not likely to produce the well defined shape consistent with a 16 Gyr isochrone (see Figure~\ref{leoi_isopa}). 

Finally, the fact that there are not obvious discontinuities in the turnoff/subgiant region suggests that the star formation in Leo~I has proceeded in a more or less continuous way, with possible changes in intensity but no big time gaps between successive bursts, through the life of the galaxy. These possible changes will be quantified, using synthetic CMDs, in Paper II.

\subsubsection{The horizontal-branch and the red-clump of core He-burning stars}\label{hb}

Core He-burning stars produce two different features in the CMD, the HB and the RC, depending on age and metallicity. Very old, very low metallicity stars distribute along the HB during the core He-burning stage. The RC is produced when the core He-burners are not so old, or more metal-rich, or both, although other factors may also play a role (see Lee 1993). The HB--RC area in Leo~I differs from those of the other dSph galaxies in the following two important ways.

First, the lack of a conspicuous HB may indicate, given the low metallicity of
the stars in the galaxy, that Leo~I has only a small fraction of very old stars. There are a number of stars at $M_I\simeq 0 $,  $(V-I)_0=0.2-0.6$ that could be
stars on the HB of an old, metal poor population, but their position is
also that of the post turn-off $\simeq$ 1 Gyr old stars (see
Figure~\ref{leoi_isoclump}). The relatively large number of these stars and the  discontinuity that can be appreciated between them and the rest of the stars in the Herszprung-Gap supports the hypothesis that HB stars may make a contribution. This possible contribution will be quantified in Paper II.  Second, the Leo~I RC is very densely populated and is much more extended in luminosity than the RC of single-age populations, with a width of as much as $\Delta I \simeq$ 1 mag. The intermediate-age LMC populous clusters with a well populated RC (see e.g. Bomans, Vallenari \& De Boer 1995) have $\Delta I$ values about a factor of two smaller. The RCs of the
other dSph galaxies with an intermediate-age population (Fornax:
Stetson et al. 1998; Carina: Hurley--Keller, Mateo \& Nemec 1998) are also much
less extended in luminosity.

The Leo~I RC is more like that observed in the CMDs of the general field of the LMC (Vallenari et al. 1996; Zaritzky, Harris \& Thompson 1997). A RC extended in luminosity is indicative of an extended SFH with a large intermediate--age component. The older stars in the core He--burning phase lie in the lower part of the observed RC, younger RC stars are brighter (Bertelli et al. 1994, their Figure~12; see also Caputo, Castellani \& Degl'Innocenti 1995). The brightest RC stars may be  $\simeq$ 1 Gyr old stars (which start the core He-burning phase in non-degenerate conditions) in their blue--loop phase. The stars scattered above the RC, (as well as the brightest  yellow stars, see subsection~\ref{yellow}), could be a few hundred Myr old in the same evolutionary phase (see Figure 1 in Aparicio et al. 1996; Gallart 1998). The RC morphology depends on the fraction of stars of different ages, and will complement the quantitative information about the SFH from the distribution of sub-giant and MS stars (Paper II).

\subsubsection{The bright  yellow stars: anomalous Cepheids?} \label{yellow}

There are a number of bright,  yellow stars in the CMD (at $-2.5 \le M_V < -1.5$ mag  and $0 \le (V-I) \le 0.6$ mag). L93 indicate that a significant fraction of these stars show signs of variability, and two of the stars in their sample were identified by Hodge \& Wright (1978) to be anomalous Cepheids\footnote{Anomalous Cepheids were first discovered in dSph galaxies, and it was demonstrated (Baade \& Swope 1961; Zinn \& Searle 1976) that they obey a period-luminosity relationship different from that of globular cluster Cepheids and classical Cepheids. The relatively large mass ($\simeq 1.5 M\odot$) estimated for them implies that they should be relatively young stars, or mass transfer binaries. Since the young age hypothesis appeared incompatible with the idea of dSph galaxies being basically Population II systems, it was suggested that anomalous Cepheids could be products of mass-transfer binary systems. Nevertheless, we know today that most dSph galaxies have a substantial amount of intermediate-age population, consistent with anomalous Cepheids being relatively young stars that, according to various authors (Gingold 1976, 1985; Hirshfeld 1980; Bono et al. 1997), after undergoing the He-flash, would evolve towards high enough effective temperatures to cross the instability strip before ascending the AGB.}. Some of them also show signs of variability in our {\it HST} data. In Figure~\ref{leoi_isoclump} however, it is shown that these stars have the magnitudes and colors expected for blue--loop stars of few hundred
Myr. This supports our previous conclusion that the brightest stars in the MS have ages similar to these.

Given their position in the CMD, it is interesting to ask whether some of the
variables found by Hodge \& Wright (1978) in Leo~I could be classical Cepheids
instead of anomalous Cepheids\footnote {Both types of variables would be double-shell burners although, taking into account the results of the authors referenced in the previous footnote, from a stellar evolution point of view the difference between them would be that the anomalous have started the He-burning in the core in degenerate conditions, while the classical are stars massive enough to have ignited He in the core under non-degenerate conditions. If the Leo~I Cepheids are indeed among the yellow stars above the RC, which are likely blue--loop stars, they would meet the evolutionary criterion to be classical Cepheids.}. From the Bertelli et al. (1994) isochrones, we can obtain the mass and luminosity of a 500 Myr blue-loop star, which would be a representative star in this position of the CMD. Such a star
would have a mass, $M\simeq 2.5 M_{\odot}$, and a luminosity, L$\simeq$
350 L$_{\odot}$.  From Eq. 8 of Chiosi et al. (1992) we calculate that
the period that corresponds to a classical Cepheid  of this mass and
metallicity is 1.2 days, which is compatible with the periods found by
Hodge \& Wright (1978), that range between 0.8 and 2.4 days.

We suggest that some of these variable stars may be similar to the
short period Cepheids in the SMC (Smith et al. 1992), i.e. classical
Cepheids in the lower extreme of mass, luminosity and period. If this
is confirmed, it would be of considerable interest in terms of
understanding the relationship between the different types of Cepheid
variables. A new wide field survey for variable stars, more accurate
and extended to a fainter magnitude limit (both to search for
Cepheids and RR Lyrae stars) would be of particular interest in the case of
Leo~I.
  
\subsubsection{The Red Giant Branch: the metallicity of Leo~I} \label{rgb}

The RGB of Leo~I is relatively blue, characteristic of a system with low metallicity. Assuming that the stars are predominantly old, with a small dispersion in age, L93 obtained a mean metallicity [Fe/H]=--2.02 $\pm$ 0.10 dex and a metallicity dispersion of $-2.3 < {\rm [Fe/H]} < -1.8$ dex. This estimate was based on the color and intrinsic dispersion in color of the RGB at $M_I=-3.5$ using a calibration based on the RGB colors of galactic globular clusters (Da Costa \& Armandroff 1990; Lee, Freedman \& Madore 1993b). For a younger mean age of about 3.5 Gyr, they estimate a slightly higher metallicity of [Fe/H]=--1.9, based on the difference in color between a 15 and a 3.5 Gyr old population according to the Revised Yale Isochrones (Green et al. 1987).  Other photometric measurements give a range in metallicity of [Fe/H]= --1.85 to --1.0 dex (see L93 and references therein). The metallicity derived from moderate resolution spectra of two giant stars by Suntzeff (1992, unpublished) is [Fe/H]$\simeq -1.8$ dex.  

Since Leo~I is clearly a highly composite stellar population with a large spread in age, the contribution to the width of the RGB from such an age range may no longer be negligible compared with the dispersion in metallicity. Therefore, an independent estimate of the age range from the MS turnoffs is relevant in the determination of the range in metallicity. In the following, we will discuss possible limits on the metallicity dispersion of Leo~I through the comparison of the RGB with the isochrones shown in Figures~\ref{leoi_isopa} through ~\ref{leoi_isoz}. As we noted in the introduction of Section~\ref{cmd}, there are some differences between the Padova and the Yale isochrones, but their positions coincide in the zone about 1 magnitude above the RC. We will use only this position in the comparisons discussed below. 

\begin{figure}
\caption[]{Z=0.0004 isochrones for 10, 4 and 1 Gyr (evolution through the RGB only) and 0.5 Gyr (full isochrone) from the Padova library, superimposed on the same data of Figure~5.}
\label{leoi_isoz}
\end{figure}

We will first check whether the whole width of the RGB can be accounted for by the dispersion in age. In subsection~\ref{ms} above, we have shown that the ages of the stars in Leo~I range from 10--15 Gyr to less than 1 Gyr. In Figure~\ref{leoi_isoz} we have superimposed Padova isochrones of Z=0.0004 and ages 10, 1 and 0.5 Gyr on the Leo~I CMD. This shows that the full width of the RGB above the RC can be accounted for by the dispersion in age alone. A similar result is obtained for a metallicity slightly lower or higher. This provides a
lower limit for the metallicity range, which could be negligible. The AGB of the 0.5 Gyr isochrone appears to be too blue compared with the stars in the corresponding area of the CMD. However, these AGBs are expected to be
poorly populated because a) stars are short lived in this phase and b) the fraction of stars younger than 1 Gyr is small, if any.   

Second, we will discuss the possible range in Z at different ages from a) the position of the RGB, taking into account the fact that isochrones of the same age are redder when they are more metal--rich and isochrones of the same metallicity are redder when they are older and b) that the extension of the blue-loops depends on metallicity: 

a) for stars of a given age, the lower limit of Z is given by the blue edge of the RGB area we are considering: isochrones of any age and Z=0.0001 have colors in the RGB above the RC within the observed range. Therefore, by means of the present comparison only, we cannot rule out the possibility that there may be stars in the galaxy with a range of ages and Z as low as Z=0.0001. The oldest stars of this metallicity would be at the blue edge of the RGB, and would be redder as they are younger. The upper limit for the metallicity of stars of a given age is given by the red edge of the RGB: for old stars, the red edge of the observed RGB implies an upper limit of Z $\le$ 0.0004 (see Figure~\ref{leoi_isoz}), since more metal rich stars would have colors redder than observed. For intermediate-age stars up to $\simeq$ 3 Gyr old we infer an upper limit of Z=0.001, and for ages $\simeq$ 3-1 Gyr old an upper limit of Z=0.004. 

b) we can use the position of the bright yellow stars to constrain Z: the fact that there are a few stars in blueward extended blue--loops implies that their metallicity is as low as Z$\leq$0.001 or even lower (Figure~\ref{leoi_isoclump}), because higher metallicity stars don't produce blueward extended blue-loops at the observed magnitude. This does not exclude the possibility that a fraction of young stars have metallicity up to Z=0.004. These upper limits are compatible with Z slowly increasing with time from Z$\simeq$ 0 to Z$\simeq$0.001--0.004, on the scale of the Padova isochrones.

In summary, we conclude that the width of the Leo~I RGB can be accounted for the dispersion of the age of its stellar population and, therefore, the metallicity dispersion could be negligible. Alternatively, considering the variation in color of the isochrones depending on both age and metallicity, we set a maximum range of metallicity of $0.0001\le {\rm Z} \le$0.001--0.004: a lower limit of Z=0.0001 is valid for any age, and the upper limit varies from Z=0.0004 to Z=0.004, increasing with time. These upper limits are quite broad; they will be better constrained, and some information on the chemical enrichment law gained, from the analysis of the CMD using synthetic CMDs in Paper II.

\section {Conclusions}\label{discus}

From the new {\it HST} data and the analysis presented in this paper, we conclude the following about the stellar populations of Leo~I:

1) The broad MS turnoff/subgiant region and the wide range in
luminosity of the RC show that star formation in Leo~I has extended
from at least $\simeq$ 10--15 Gyr ago to less than 1 Gyr ago.  A lack
of  obvious discontinuities in the MS turnoff/subgiant region suggests that star
formation proceeded in a more or less continuous way in the central
part of the galaxy, with possible intensity variations over time, but no big time gaps between successive bursts, through the life of the galaxy.

2) A conspicuous HB is not seen in the CMD. Given the low metallicity of the galaxy, this reasonably  implies that the fraction of stars older than $\simeq$ 10 Gyr is small, and indicates that the beginning of a substantial amount of star formation may have been delayed in Leo~I in comparison to the other dSph galaxies.  It is unclear from the analysis presented in this paper whether Leo~I contains any stars as old as the Milky Way globular clusters.

3) There are a number of bright, yellow stars in the same area of the
CMD where anomalous Cepheids have been found in Leo I. These stars also
have the color and magnitude expected for the blue-loops of low
metallicity, few--hundred Myr old stars. We argue that some of these
stars may be classical Cepheids in the lower extreme of mass,
luminosity and period.

4)The evidence that the stars in Leo~I have a range in age complicates the determination of limits to the metallicity range based on the width of the RGB. In one extreme, if the width of the Leo~I RGB is atributted to the dispersion of the age of its stellar population alone, the metallicity dispersion could be negligible. Alternatively, considering the variation in color of the isochrones depending on both age and metallicity, we set a maximum range of metallicity of $0.0001\le {\rm Z} \le$0.001--0.004: a lower limit of Z=0.0001 is valid for any age, and the (broad) upper limit varies from Z=0.0004 to Z=0.004, increasing with time. 

In summary, Leo~I has unique characteristics among Local Group
galaxies. Due to its morphology and its lack of detectable quantities of
HI, it can be classified as a dSph galaxy. But it appears to have the youngest
stellar population among them, both because  it is the only dSph lacking a conspicuous old population, and because it seems to have a larger fraction of
intermediate-age and young population than other dSph. The
star formation seems to have proceeded until almost the present time,
without evidence of intense, distinct bursts of star formation.

Important questions about Leo~I still remain.  An analysis of the data
using synthetic CMDs will give quantitative information about the
strength of the star formation at different epochs. Further
observations are needed to characterize the variable-star population in
Leo~I, and in particular, to search for RR Lyrae variable stars. This
will address the issue of the existence or not of a very old stellar
population in Leo~I. It would be interesting to check for variations of
the star formation across the galaxy and to determine whether the HB
is also missing in the outer parts of Leo~I.

Answering these questions is important not only to understand the
formation and evolution of Leo~I, but also in relation to general
questions about the epoch of galaxy formation and the evolution of
galaxies of different morphological types. The determination of the
strength of the star formation in Leo~I at different epochs is
important to assess whether it is possible that during intervals of high star
formation activity, Leo~I would have been as bright as the faint blue
galaxies observed at intermediate redshift. In addition, the duration
of such a major event of star formation may be important in explaining
the number counts of faint blue galaxies.

\acknowledgments

We want to thank Allan Sandage for many very useful discussions and a
careful reading of the manuscript. We thank also Nancy B. Silbermann, 
Shoko Sakai and Rebecca Bernstein for their help through the various stages of the {\it HST} data reduction. Support for this work was provided by NASA grant
GO-5350-03-93A from the Space Telescope Science Institute, which is
operated by the Association of Universities for Research in Astronomy
Inc. under NASA contract NASA5--26555.  C.G. also acknowledges
financial support from a Small Research Grant from NASA administered 
by the AAS and a Theodore Dunham Jr. Grant for Research in Astronomy. A.A. thanks the Carnegie Observatories for their hospitality. A.A. is supported by the Ministry of Education and Culture of the Kingdom of Spain, by the University of La Laguna and by the IAC (grant PB3/94). M.G.L is supported by the academic research fund of Ministry of Education, Republic of Korea, BSRI-97-5411. The Digitized Sky Surveys were produced at the Space Telescope Science Institute under U.S. Government grant NAG W-2166. The images of these surveys are based on photographic data obtained using the Oschin Schmidt Telescope on Palomar Mountain and the UK Schmidt Telescope.

\newpage

\end{document}